\documentclass[prl,showpacs, twocolumn,floatfix]{revtex4}

\usepackage{graphicx}

\begin{document}
\title{Observation of Antinormally Ordered Hanbury-Brown$-$Twiss Correlations}
\author{Koji Usami$^{1,2}$} \email{usami@frl.cl.nec.co.jp}
\author{Yoshihiro Nambu$^{2,3}$}
\author{Bao-Sen Shi$^{4}$}
\author{Akihisa Tomita$^{1,2,3,4}$}
\author{Kazuo Nakamura$^{1,2,3}$}  
\affiliation{$^1$Department of Material Science and Engineering, Tokyo Institute of Technology, Yokohama 226-0026, Japan \\
$^2$CREST, Japan Science and Technology Agency (JST), Tokyo 150-0002, Japan \\
$^3$Fundamental Research Laboratories, NEC, Tsukuba 305-8501, Japan \\
$^4$ERATO, Japan Science and Technology Agency (JST), Tokyo 113-0033, Japan}

\date{\today }

\begin{abstract}
We have measured antinormally ordered Hanbury-Brown$-$Twiss correlations for coherent states of electromagnetic field by using stimulated parametric down-conversion process. Photons were detected by stimulated emission, rather than by absorption, so that the detection responded not only to actual photons but also to zero-point fluctuations via spontaneous emission. The observed correlations were distinct from normally ordered ones as they showed excess positive correlations, i.e., photon bunching effects, which arose from the thermal nature of zero-point fluctuations. 
\end{abstract}


\maketitle

Since Planck's quantization hypothesis of electromagnetic field and Einstein's photoelectric theory appeared, the quantum nature of electromagnetic field has been intensively studied~\cite{Loudon}. The normally ordered photodetection theory of Glauber~\cite{Glauber1963} played a central role in these studies~\cite{Loudon}. This theory provided a formal explanation of the Hanbury-Brown$-$Twiss (HBT) type correlation measurements~\cite{HBT1956}, in which the electromagnetic fields were detected at two separated space-time points. The normal ordering reflects the fact that the electromagnetic field is detected by an absorption process. Thus, such a photodetection is insensitive to zero-point fluctuations because the photodetection probability for the vacuum state is zero, i.e., $\langle 0|\hat{a}^{\dagger}\hat{a}|0 \rangle=0$. This insensitivity is the very reason why Planck's spectrum of black-body radiation is always convergent, even if the electromagnetic energy caused by zero-point fluctuations is divergent~\cite{GZ}. It is also the reason why pieces of information on the field, more precisely, the vacuum components of the field's density matrix, are lost during the detection process and thus the initial density matrix cannot be \textit{logically reversible}~\cite{Ueda}, that is, cannot be calculated from the post-detection density matrix and the readout of the detection.

When photons are detected by stimulated emission, however, antinormally ordered photodetection can be realized~\cite{Mandel1966}. Since the detection responds not only to actual photons but also to zero-point fluctuations via spontaneous emissions, the system's information can in principle be conserved during this photodetection process~\cite{Ueda}. The photon-counting statistics of the detection is then distinct from that of the standard normally ordered photodetection especially in the region where the average photon-number in the concerned modes is small~\cite{Mandel1966}. Although the emission-based antinormally ordered photodetector was originally proposed by Bloembergen, as the \textit{quantum counter}, as early as 1959~\cite{Bloembergen1959}, no report on the experimental realization exists till date to the best of our knowledge.

As a demonstration of the antinormally ordered photodetection, we realized the quantum counters by using stimulated parametric down-conversion process and measured the HBT correlations for coherent states in antinormal order. Here, we utilized an ultrashort pulsed laser as a pump field for the parametric process to obtain large nonlinear response of the crystal and to overcome the slow response time of the detectors~\cite{Koashi,ORW1999}. The observed correlations deviated from the standard normally ordered ones as they exhibited excess positive correlations, i.e., photon bunching effects. The deviation can be attributed to zero-point fluctuations, which are known to possess thermal characteristics~\cite{thermalT,ORW1999}.

\begin{figure*}
\begin{center}
\includegraphics[width=0.9\linewidth]{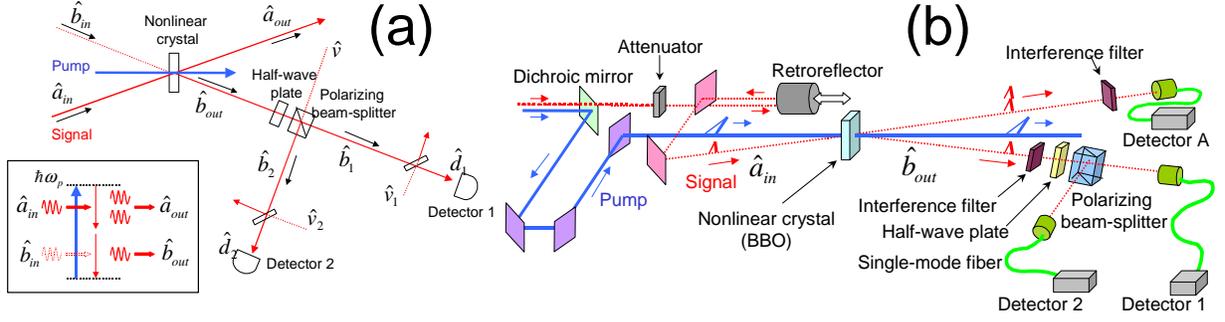}
\caption{(a) Schematic illustration of the antinormally ordered HBT correlator based on stimulated parametric down-conversion. (b) Experimental setup (the mode-locked Ti:Sapphire laser and the frequency doubling crystal are not shown but placed at the left side of the figure). The correlations for the signal field, $\hat{a_{in}}$, are acquired by counting the delayed-coincidental photodetection events of fields $\hat{b_{1}}$ and $\hat{b_{2}}$ with detectors~1 and 2.}
\label{fig:setup2}
\end{center}
\end{figure*}

To clarify how to measure the antinormally ordered HBT correlation, we pursue the time evolution of a single-mode annihilation operator, $\hat{a_{in}}$, as the signal field. A schematic illustration of the correlator is shown in Fig.~\ref{fig:setup2}~(a). First, operator $\hat{a_{in}}$ is coupled with an operator, $\hat{b_{in}}$, via parametric interaction with the pump field and evolved into $\hat{a_{out}}=\hat{a_{in}} \cosh [s L]-\hat{b_{in}}^{\dagger} e^{i \vartheta} \sinh [s L]$; while operator $\hat{b_{in}}$ becomes $\hat{b_{out}}=\hat{b_{in}} \cosh [s L]-\hat{a_{in}}^{\dagger} e^{i \vartheta} \sinh [s L]$ under a perfectly phase-matched condition~\cite{Loudon}. Here, $L$ is the length of the crystal, and several parameters in the interaction, such as the intensity of the pump field and the second-order nonlinear susceptibility of the crystal are included in parameters $s$ and $\vartheta$. An energy diagram of this process is shown in the inset of Fig.~\ref{fig:setup2}~(a). Next, by dividing the field represented by $\hat{b_{out}}$ into fields $\hat{b_{1}}$ and $\hat{b_{2}}$ with a half-wave plate and a polarizing beam-splitter, the standard HBT interferometer for field $\hat{b_{out}}$ is formed as shown in Fig.~\ref{fig:setup2}~(a). Here, the imperfect quantum efficiencies of the photodetectors and several optical losses are taken into account and modeled by introducing auxiliary beam-splitters with vacuum fields $\hat{v_{1}}$ and $\hat{v_{2}}$~\cite{Loudon}. Output fields $\hat{d_{1}}$ and $\hat{d_{2}}$ of Fig.~\ref{fig:setup2}~(a) are then written as $\hat{d_{1}} = \sqrt{\eta_{1}}\ (\mathcal{T} \ \hat{b_{out}}+\mathcal{R} \ \hat{v}) + i \sqrt{1-\eta_{1}} \ \hat{v_{1}}$ and $\hat{d_{2}} = \sqrt{\eta_{2}}\ (\mathcal{R} \ \hat{b_{out}}+\mathcal{T} \ \hat{v}) + i \sqrt{1-\eta_{2}} \ \hat{v_{2}}$, respectively. Here, $\eta_{1}$ and $\eta_{2}$ are the total photodetection efficiencies for detectors~1 and 2, respectively; $\mathcal{R}$ and $\mathcal{T}$ denote the reflection and transmission coefficients at the beam-splitter, respectively, and can be varied with the half-wave plate and the polarizing beam-splitter; and $\hat{v}$ is an auxiliary vacuum field introduced from the empty port of the polarizing beam-splitter. Since the modes relevant to operators $\hat{b_{in}}$, $\hat{v}$, $\hat{v_{1}}$, and $\hat{v_{2}}$ are initially vacua, the surviving contribution to the number of the normally ordered photodetection events for fields $\hat{d_{1}}$ and $\hat{d_{2}}$ coincides with that of the antinormally ordered photodetection for field $\hat{a_{in}}$ up to a constant factor, i.e., $\langle \hat{n}_{d_{1}} \rangle \equiv \langle \hat{d_{1}}^{\dagger} \hat{d_{1}} \rangle = \eta_{1}|\mathcal{T}|^{2}\sinh^{2} [s L] \langle \hat{a_{in}} \hat{a_{in}}^{\dagger} \rangle$ and $\langle \hat{n}_{d_{2}} \rangle \equiv \langle \hat{d_{2}}^{\dagger} \hat{d_{2}} \rangle = \eta_{2}|\mathcal{R}|^{2}\sinh^{2} [s L] \langle \hat{a_{in}} \hat{a_{in}}^{\dagger} \rangle$, respectively. Here, the angle brackets indicate quantum-mechanical expectation values. Furthermore, the number of coincidental photodetection events of fields $\hat{d_{1}}$ and $\hat{d_{2}}$ results in $\langle \hat{n}_{d_{1}} \hat{n}_{d_{2}} \rangle \equiv \langle \hat{d_{1}}^{\dagger} \hat{d_{1}} \hat{d_{2}}^{\dagger} \hat{d_{2}} \rangle = \eta_{1}\eta_{2}|\mathcal{T}|^{2}|\mathcal{R}|^{2}\sinh^{4} [s L]  \langle \hat{a_{in}} \hat{a_{in}} \hat{a_{in}}^{\dagger} \hat{a_{in}}^{\dagger} \rangle$, where we use the commutation relation for each operator and relations $\mathcal{TR^{*}}=-\mathcal{RT^{*}}$ and $\cosh^{2} [s]-1=\sinh^{2} [s]$. Thus, the surviving contribution to the coincidences turns out to be made only by operators $\hat{a_{in}}$ and $\hat{a_{in}}^{\dagger}$ in antinormal order. Consequently, we can evaluate an antinormally ordered HBT correlation for field $\hat{a_{in}}$ as follows:
\begin{equation}
g^{(2)}_{1,2} \equiv \frac{\langle \hat{n}_{d_{1}} \hat{n}_{d_{2}} \rangle}{\langle \hat{n}_{d_{1}} \rangle \langle \hat{n}_{d_{2}} \rangle} = \frac{\langle \hat{a_{in}} \hat{a_{in}} \hat{a_{in}}^{\dagger} \hat{a_{in}}^{\dagger} \rangle}{\langle \hat{a_{in}} \hat{a_{in}}^{\dagger} \rangle \langle \hat{a_{in}} \hat{a_{in}}^{\dagger} \rangle} \equiv g^{(2[A])}.  \label{eq:ANHBT}
\end{equation}
Note that Eq.~(\ref{eq:ANHBT}) holds regardless of the splitting ratio at the beam-splitter, the quantum efficiencies of the detectors, and the optical losses.

To analyze the correlation more realistically, the detection process should be treated with time-dependent and continuous-mode field operators~\cite{Loudon}. In this treatment, the antinormally ordered HBT correlation, Eq.~(\ref{eq:ANHBT}), becomes time dependent. Moreover, since the response-time jitter of the detector is larger than the pump pulse duration (the duration of the parametric interaction) but smaller than the time interval between two successive pulses, the relevant information on the time dependence is embodied in the integrated number of delayed coincidences over the response-time jitter of the detectors~\cite{Koashi,ORW1999}. Thus, what we should measure to evaluate the antinormally ordered HBT correlation is the correlation of the pulses of $m$th neighbor:
\begin{equation}
g^{(2[A])}_{\ m} \equiv \frac{\langle \int_{t}^{t+T}\!\!\! dt' \int_{t+m\tau}^{t+m\tau+T} \!\!\! dt''  \hat{d_{1}}^{\dagger}(t') \hat{d_{1}}(t') \hat{d_{2}}^{\dagger}(t'') \hat{d_{2}}(t'') \rangle}{\langle \int_{t}^{t+T}\!\!\! dt' \hat{d_{1}}^{\dagger}(t') \hat{d_{1}}(t') \rangle \langle \int_{t+m\tau}^{t+m\tau+T} \!\!\! dt'' \hat{d_{2}}^{\dagger}(t'') \hat{d_{2}}(t'') \rangle}, \label{eq:m_ANHBT}
\end{equation}
where $t$ and $t+m\tau$ are the initial time of the integration for detectors~1 and 2, respectively, $T$ is the duration of integration, and $\tau$ corresponds to the time interval between two successive pump pulses. Here, since the coherence time of the down-converted field is far shorter than the pump pulse interval, we have $g^{(2[A])}_{\ m}=1$ for $m\neq0$, thus we only focus on the value, $g^{(2[A])}_{\ 0}$. Note that the duration, $T$, can be approximated by infinity because the response-time jitter of the detector is far longer than any other relevant time scale. By taking account of the pump spectrum and the bandwidth of the interference filter~\cite{ORW1999,HSB1990}, the correlation, $g^{(2[A])}_{\ 0}$ of Eq.~(\ref{eq:m_ANHBT}), for a coherent state becomes~\cite{Usami}
\begin{equation}
g^{(2[A])}_{\ 0}=1+\alpha \big[\, \frac{1}{\langle \hat{n} \rangle +1}+\frac{\langle \hat{n} \rangle}{[\,\langle \hat{n} \rangle +1\,]^{2}} \,\big], \label{eq:ANHBT_emC}
\end{equation}
where $\langle \hat{n} \rangle \equiv \langle \hat{a_{in}}^{\dagger}\hat{a_{in}} \rangle$ is the average photon number of the signal field. The parameter, $\alpha$, represents the indistinguishability of two emitted photons, which are responsible for a coincidental count. Here, the value, $\alpha=1$, corresponds to the case that two emitted photons are completely indistinguishable.

\begin{figure}
\begin{center}
\includegraphics[width=\linewidth]{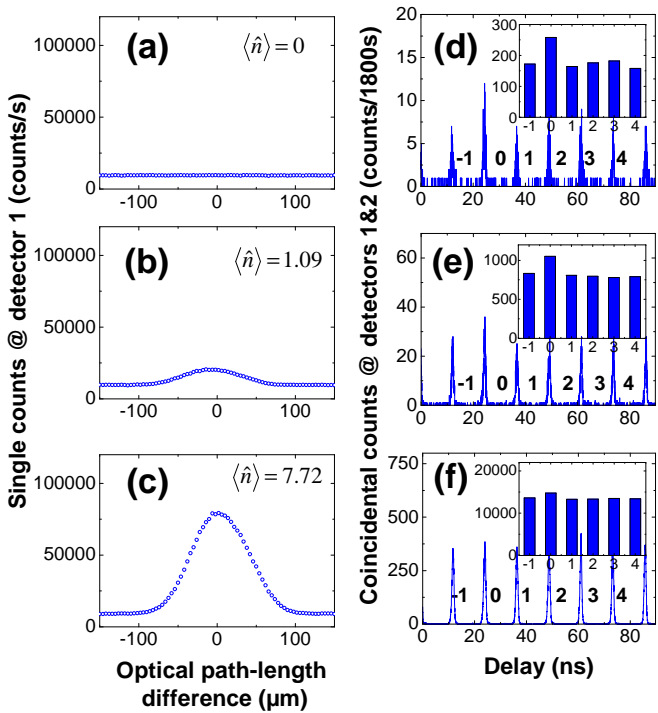}
\caption{Experimental results. (a)-(c):~The number of single counts in detector~1 as a function of the optical path-length difference between the pump field and the input signal field, $\hat{a_{in}}$, whose average photon number $\langle \hat{n} \rangle$ is (a) $0$ (vacuum), (b) $1.09$, and (c) $7.72$. (d)-(f):~The number of delayed-coincidental counts in detectors 1 and 2. Here, the coherent field with the average photon number indicated in each of the figure on the left was used as input signal. The insets indicate the accumulated number of coincidences within 3~ns.}
\label{fig:results}
\end{center}
\end{figure}

A rough sketch of our experimental setup is shown in Fig.~\ref{fig:setup2}~(b). A parametric down-converter was formed with a beta-barium borate (BBO) crystal (thickness: 2~mm) and a pulsed pump field (wavelength:~399~nm, average power:~195~mW, pulse duration:~100~fs, and repetition rate:~82~MHz) from the second harmonic of a mode-locked Ti:Sapphire laser (Spectra-Physics Tsunami). The crystal was arranged to be type-I phase-matched so that vertically polarized frequency-degenerate down-converted photons (wavelength:~798~nm) were spontaneously emitted as the spatially-nondegenerate fields, $\hat{a_{out}}$ and $\hat{b_{out}}$, separated by $\pm2.54^{\circ}$ with respect to the pump. A heavily attenuated coherent state (a laser output with the fundamental wavelength (798~nm)) was used as a signal field represented by $\hat{a_{in}}$. Its average photon number can be varied by the attenuator in Fig.~\ref{fig:setup2}~(b). The signal field, $\hat{a_{in}}$, was amplified via stimulated parametric down-conversion when the optical paths of the signal and the pump fields were properly adjusted with the retroreflector and the mirrors shown in Fig.~\ref{fig:setup2}~(b). One of the stimulated fields, $\hat{b_{out}}$, was further split into fields $\hat{b_{1}}$ and $\hat{b_{2}}$ with a half-wave plate and a polarizing beam-splitter as shown in Fig.~\ref{fig:setup2}~(b). After propagating about 800~mm from the crystal and passing through an interference filter (bandwidth: 5.0~nm) each field was coupled into a single-mode fiber, which acted as a spatial filter, and then detected by Si-avalanche photodiodes (Perkin Elmer SPCM-AQR-14; detection efficiency: 55~\%, response-time jitter: 350~ps, and dark-count rate: 100~s$^{-1}$). Electric pulses produced by the detectors were processed by digital-logic gates (Kaizu Works KN470), and the number of coincidences (threefold coincidences) in two (three) detectors as well as single counts in each detector were counted with pulse counters (Stanford Research System SR620) for aligning the setup. The pulses from detectors~1 and 2 were also fed into a time-interval analyzer (Yokogawa TA520; time resolution: 25~ps) for measuring their delayed coincidences.

Figures~\ref{fig:results}(a)-(c) show the number of single counts in detector~1 (during 1~s) as a function of the optical path-length difference between the pump field and the signal field, $\hat{a_{in}}$, where average photon numbers $\langle \hat{n} \rangle$ of the signal fields for (a), (b), and (c), differ each other. Here, we set the half-wave plate in Fig.~\ref{fig:setup2}(b) to be $|\mathcal{T}|^{2}=1$ and $|\mathcal{R}|^{2}=0$. The counts were enhanced by the stimulated emission within the area where the optical path-length difference was small, i.e., two pulses overlapped. On the other hand, the constant background counts, which were independent of the optical path-length difference, were attributed to the spontaneous emissions. The maximally enhanced counts due to the stimulations are given by $\langle \hat{n}_{d_{1}} \rangle = \eta_{1}|\mathcal{T}|^{2}\sinh^{2} [s] (\langle \hat{n} \rangle +1)$, whereas the counts without any stimulations are given by $\langle \hat{n}_{d_{1}} \rangle = \eta_{1}|\mathcal{T}|^{2}\sinh^{2} [s]$. Thus, by comparing these counts the signal field's average photon number, $\langle \hat{n} \rangle$, can be quantitatively evaluated as (a) $0$ (vacuum), (b) $1.09$, and (c) $7.72$, respectively.

Figures~\ref{fig:results}(d)-(f) show the number of coincidental counts in detectors 1 and 2 (during 1800~s) recorded for various time delays (25-ps time bin). Here, the coherent field with the average photon number indicated in each of the figure on the left was used as input signal, and the half-wave plate was set to be $|\mathcal{T}|^{2}=1/2$ and $|\mathcal{R}|^{2}=1/2$. The peaks of the coincidences at 12.2-ns intervals corresponded to the 82-MHz repetition rate of the mode-locked laser, and the width of each peak was dominantly determined by the 350-ps response-time jitter of the detectors. As mentioned before, the relevant information on the correlation should be extracted after accumulating the number of coincidences within the width of each peak. The insets in Figs.~\ref{fig:results}(d)-(f) indicate the accumulated number of coincidences within 3~ns, which is sufficiently larger than the response-time jitter of the detector. The second peaks in Figs.~\ref{fig:results}(d)-(f) correspond to the coincidences at zero time delay ($m=0$), where two photons produced by the same pump pulse are responsible for the coincidences, and they are larger than any other peak. 

\begin{figure}
\begin{center}
\includegraphics[width=0.9\linewidth]{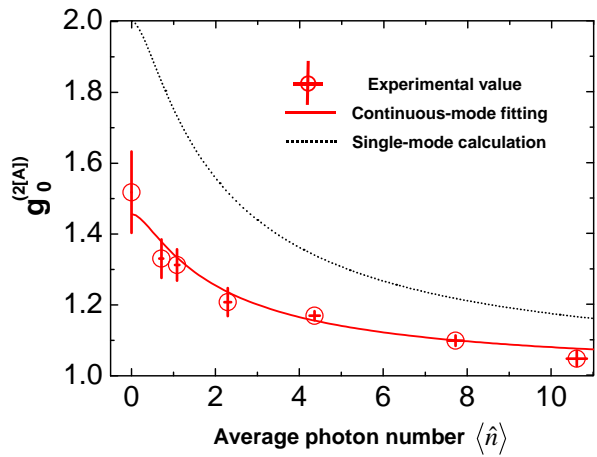}
\caption{Normalized antinormally ordered HBT-type correlation $g^{(2[A])}_{\ 0}$. The open circles indicate the experimentally evaluated correlations with error bars for both axes. The dotted-line curve is the theoretical prediction obtained by the single-mode calculation. The full-line curve was obtained by a $\chi^{2}$ fitting of the data to the prediction with continuous-mode treatment.}
\label{fig:g2A}
\end{center}
\end{figure}

The antinormally ordered HBT correlations, $g^{(2[A])}_{\ 0}$, were calculated as follows. As mentioned before, unless the delay $m\tau$ in Eq.~(\ref{eq:m_ANHBT}) is zero, $g^{(2[A])}_{\ m}$ can be regarded as one, i.e, no correlation survives. Therefore, the accumulated number of coincidences in the second peak ($m=0$) normalized by that of the non-zero delay peak ($m\neq0$) can be considered as the correlation, $g^{(2[A])}_{\ 0}$. To evaluate the variance as well, we used five peaks ($m=-1,1,2,3,4$) at non-zero delay to determine the normalization factor. The averages of the correlations for seven input fields (the average photon numbers are 0, 0.71, 1.09, 2.29, 4.36, 7.72, and 10.61) are shown in Fig.~\ref{fig:g2A} as the open circles with vertical error bars (three standard deviations). Each horizontal error bar (three standard deviations) was evaluated by assuming that the number of photodetections followed the Poissonian distribution. As opposed to the normally ordered HBT correlations for coherent states, the results show the photon bunching effects ($g^{(2[A])}_{\ 0} \ge g^{(2[A])}_{\ m \ne 0}=1$), which were triggered by the influence of zero-point fluctuations~\cite{thermalT,ORW1999}. The bunching effect becomes inconspicuous as the average photon number of the input field increases and approaches the normally ordered value, $g^{(2)}_{\ 0}=1$~\cite{Mandel1966}. In this sense, this discrepancy can be viewed as a purely quantum effect. From another point of view, these bunching effects can be ascribed to the fact that a coherent state is not an eigenstate of a creation operator, i.e., a measurement operator of the antinormally ordered photodetection. Thus, the coherent properties of coherent states no longer hold in the antinormally ordered HBT correlations.

The dotted-line curve in Fig.~\ref{fig:g2A} indicates value $g^{(2[A])}_{\ 0}$ of Eq.~(\ref{eq:ANHBT_emC}) with $\alpha=1$, that is, the simple prediction obtained by Eq.~(\ref{eq:ANHBT}), where all the relevant fields are treated as single-mode states. The full-line curve in Fig.~\ref{fig:g2A} indicates the value of $g^{(2[A])}_{\ 0}$ with $\alpha=0.45$, which is obtained by applying a $\chi^{2}$ fitting with $\alpha$ as a fitting parameter, i.e., by finding the minimum value of $s=\sum_{i=1}^{7}\frac{1}{\sigma_{y_{i}}^{2}}(y_{i}-[1+\alpha (\frac{1}{\langle \hat{n} \rangle_{i} +1}+\frac{\langle \hat{n} \rangle_{i}}{(\langle \hat{n} \rangle_{i} +1)^{2}})])^{2}$. Here, $\langle \hat{n} \rangle_{i}$, $y_{i}$, and $\sigma_{y_{i}}$ are the average photon number, the correlation, and the standard deviation of the correlation for $i$th experimental value in Fig.~\ref{fig:g2A}, respectively, and the variances of the horizontal axis are neglected. To take a quantitative look at how good the fitting is, we executed a $\chi^{2}$ test of goodness-of-fit. The statistical distribution of value $s$ is supposed to obey the $\chi^{2}$ distribution with six degrees of freedom (seven data $-$ one unknown parameter), which has value $\chi^{2}(6)=12.6$ at the upper 5\% point. On the other hand, with the best fitting ($\alpha=0.45$), the value, $s$, reaches 10.2, which is well below 12.6, thus, we can conclude that the experimental results are in good agreement with the continuous-mode analysis of the detection process. A detailed analysis~\cite{Usami} shows that the resultant reduction of correlation, i.e., $\alpha=0.45$, can be mainly attributed to the imperfect erasing of the time-stamp information of the down-conversion processes (reduces $\alpha$ to about 0.9), and the spatial-mode mismatching (reduces $\alpha$ further to about 0.55).

In conclusion, to demonstrate antinormally ordered photodetection, the HBT correlations for coherent states have been measured in antinormal order by using stimulated parametric down-conversion. Since the measurement operator was no longer an annihilation operator, but rather a creation operator, even the coherent states exhibited the photon bunching effects. The emission-based antinormally ordered photodetection may provide an interesting alternative for monitoring quantum systems owing to its sensitivity to zero-point fluctuations. 

\begin{acknowledgments}
We thank Y.~Tsuda, S.~Kono, H.~Kosaka, S.~Ishizaka, T.~Hiroshima, T.~Kimura, J.~Ushida, A.~V.~Gopal, Y.~Nakamura, T.~Yamamoto, and M.~Kozuma for fruitful discussions. We also acknowledge the valuable input by G.~Weihs on the fiber-based experiment and the stimulating comment of M.~Ueda on the prospects of the logically reversible measurement.
\end{acknowledgments}

\end{document}